\documentclass[sigconf]{acmart}
\usepackage{booktabs, subfigure, threeparttable} 
\usepackage{epsfig}
\usepackage{amssymb}
\usepackage{multirow}
\usepackage{booktabs}
\usepackage{tabularx}
\usepackage{graphics}
\usepackage{wrapfig}
\usepackage{subfigure}
\usepackage{caption}
\usepackage{lipsum}
\usepackage[linesnumbered,ruled]{algorithm2e}
\usepackage{flushend}
\usepackage{graphicx}
\usepackage{epstopdf}

\setcopyright{acmcopyright}

\copyrightyear{2020}
\acmYear{2020}
\setcopyright{iw3c2w3}
\acmConference[WWW '20]{Proceedings of The Web Conference 2020}{April 20--24, 2020}{Taipei, Taiwan}
\acmBooktitle{Proceedings of The Web Conference 2020 (WWW '20), April 20--24, 2020, Taipei, Taiwan}
\acmPrice{}
\acmDOI{10.1145/3366423.3380025}
\acmISBN{978-1-4503-7023-3/20/04}

\begin{document}
\title{To be Tough or Soft: Measuring the Impact of Counter-Ad-blocking Strategies on User Engagement}

\author{Shuai Zhao}
\affiliation{%
  \institution{New Jersey Institute of Technology}
}
\email{sz255@njit.edu}

\author{Achir Kalra}
\affiliation{%
  \institution{Forbes Media}
}
\email{akalra@forbes.com}

\author{Cristian Borcea}
\affiliation{%
  \institution{New Jersey Institute of Technology}
}
\email{borcea@njit.edu}
\author{Yi Chen}
\affiliation{%
  \institution{New Jersey Institute of Technology}
}
\email{yi.chen@njit.edu}

\newcommand{\sz}[1]{\textcolor{black}{#1}}
\newcommand{\yc}[1]{\textcolor{orange}{#1}}
\newcommand{\cb}[1]{\textcolor{green}{#1}}

\begin{abstract}
The fast growing ad-blocker usage results in large revenue decrease for ad-supported online websites. Facing this problem, many online publishers choose either to cooperate with ad-blocker software companies to show \emph{acceptable ads} or to build a \emph{wall} that requires users to whitelist the site for content access. However, there is lack of studies on the impact of these two counter-ad-blocking strategies on user behaviors. To address this issue, we conduct a randomized field experiment on the website of Forbes Media, a major US media publisher. The ad-blocker users are divided into a treatment group, which receives the wall strategy, and a control group, which receives the acceptable ads strategy. We utilize the difference-in-differences method to estimate the causal effects. Our study shows that the wall strategy has an overall negative impact on user engagements. However, it has no statistically significant effect on high-engaged users as they would view the pages no matter what strategy is used. It has a big impact on low-engaged users, who have no loyalty to the site.  Our study also shows that revisiting behavior decreases over time, but the  ratio of session whitelisting increases over time as the remaining users have relatively high loyalty and high engagement. The paper concludes with  discussions of managerial insights for publishers when determining counter-ad-blocking strategies.
\end{abstract}

\copyrightyear{2020}
\acmYear{2020}
\acmConference[WWW '20]{Proceedings of The Web Conference 2020}{April 20--24,
2020}{Taipei, Taiwan}
\acmBooktitle{Proceedings of The Web Conference 2020 (WWW '20), April 20--24, 2020,
Taipei, Taiwan}
\acmPrice{}
\acmDOI{10.1145/3366423.3380025}
\acmISBN{978-1-4503-7023-3/20/04}

\ccsdesc[500]{General and reference~Empirical studies}
\ccsdesc[500]{Information systems~Online advertising}
\ccsdesc[500]{Human-centered computing~User studies}

\keywords{Ad Blocking, Randomized Field Experiment, Difference-In-Differences, User Studies, Online Advertising}

\maketitle

\section{Introduction}  \label{sec:intro}

An ad blocker is a tool, most likely a browser plugin, to remove ads while a user is reading online content. The broad usage of ad blockers has a big impact to the ad-supported web publishing system. Web publishers provide  content for free, instead they gain revenue from  digital advertising, which contributes 333.25 billion US dollars  in 2019~\cite{ad_2019}. With the increasing usage of ad blockers,  it  is expected that online publishers will lose revenue of 35 billion US dollars   world-wide by 2020, and the loss has a steady increase at 30\% per year~\cite{adblocker_2020}. Without sufficient revenue, publishers cannot afford to generate high-quality free content, which will ultimately hurt online users' interests.

In response, more and more online publishers (e.g., Wired, Forbes, AdAge, Digiday, Los Angeles Daily News) launched their counter-ad-blocking methods~\cite{zhao2017ad}. Rafique et al. found that counter-ad-blocking scripts were used by 16.3\% of the 1,000 most popular domains~\cite{rafique2016s}. Currently, there are two popular methods of counter-ad-blocking used by publishers. The first is the {\em tough} ``whitelist-or-leave'' strategy, and the second is the {\em soft} acceptable ads exchange (AAX) strategy. The ``whitelist-or-leave'' strategy works like a wall (called Wall strategy as well). When an ad blocker is detected, a publisher's website pops up a message requesting the user to turn off or pause the ad blocker, i.e., whitelist the publisher's website. If a user rejects the request, she is forbidden to access the content that she intends to view. The soft AAX strategy shows users acceptable ads, agreed upon with the ad blocking companies, which appear in the page even when an ad blocker is active. Acceptable ads are generally less annoying ads, such as text ads instead of video ads, and also fewer in number. 

Despite of the importance of the ad blocking problem, there are few studies on it. Existing work ~\cite{pujol2015annoyed,iqbal2017ad}  focused on techniques and mechanisms  for counter-ad-blocking, but  not  the effect of different counter-ad-blocking strategies on users' engagement. The work in~\cite{sinha2017anti}, on the other hand, has studied the effect of such strategies. However, it compared the Wall strategy with the ads-free strategy under a retrospective quasi-experiment setting. It is not realistic for publishers to provide free content with no ads. We have different experiment setting and the goal of our study is to understand in-depth the differences of effect between the Wall strategy and the AAX strategy on user engagement, both of which are actively used by online publishers. Specifically, we want to address the following research questions:

\begin{itemize}
  \item \textbf{RQ1}: What is the overall effect of the Wall strategy on user engagement compared to the AAX strategy? Furthermore, what is the effect if an ad blocker user chooses to whitelist? 
  \item \textbf{RQ2}: How does the effect differ for user groups with different characteristics?
  \item \textbf{RQ3}: What is the longer-term effect of the Wall strategy? How would that differ from the short-term effect?
 \end{itemize}

\textbf{Contributions.} To the best of our knowledge, this is the first study to compare the two most commonly used counter-ad-blocking strategies on the web. Our work contributes empirical evidences for understanding the different impact of these two strategies on user engagement. Our study shows that the Wall strategy has an overall negative impact on user engagements. 
It has no statistically significant effect on highly-engaged users because they would view the pages no matter what strategy is used. On the other hand, it  has a big impact on low-engaged users, who have no loyalty to the site, especially in terms of reduced number of page views. Our long-term study finds that revisiting behavior decreases over time, but the ratio of session whitelisting increases over time because the remaining users have relatively high loyalty and high engagement. Although our work uses user behavior data from  one publisher, given that the datasets and settings are common to most of publishers, we expect our findings  generalizable to most online publishers.

\section{Experiment Design} \label{sec:design}

All users in our study are ad blocker users. The experiment ran for a period of two and half months in 2018, from August 13th to October 22th. On September 13th of 2018, the Wall strategy started. Before that, all users received the AAX strategy. \sz{For each incoming user, we randomly assigned her to either control or treatment group and used cookies to track the user over time. It is noted that one inevitable limitation of using cookie is that we will not be able to identify a user if she deletes her cookie. However, identifying users beyond using cookies would violate user privacy regulations (e.g., GDPR~\footnote{General Data Protection Regulation,  https://ec.europa.eu/info/law/law-topic/data-protection\_en}).} There were no other significant changes to the website during the experiment, which avoids confounding or extraneous factors brought by the publisher.

\begin{figure}[h!]
\centering
\epsfig{file=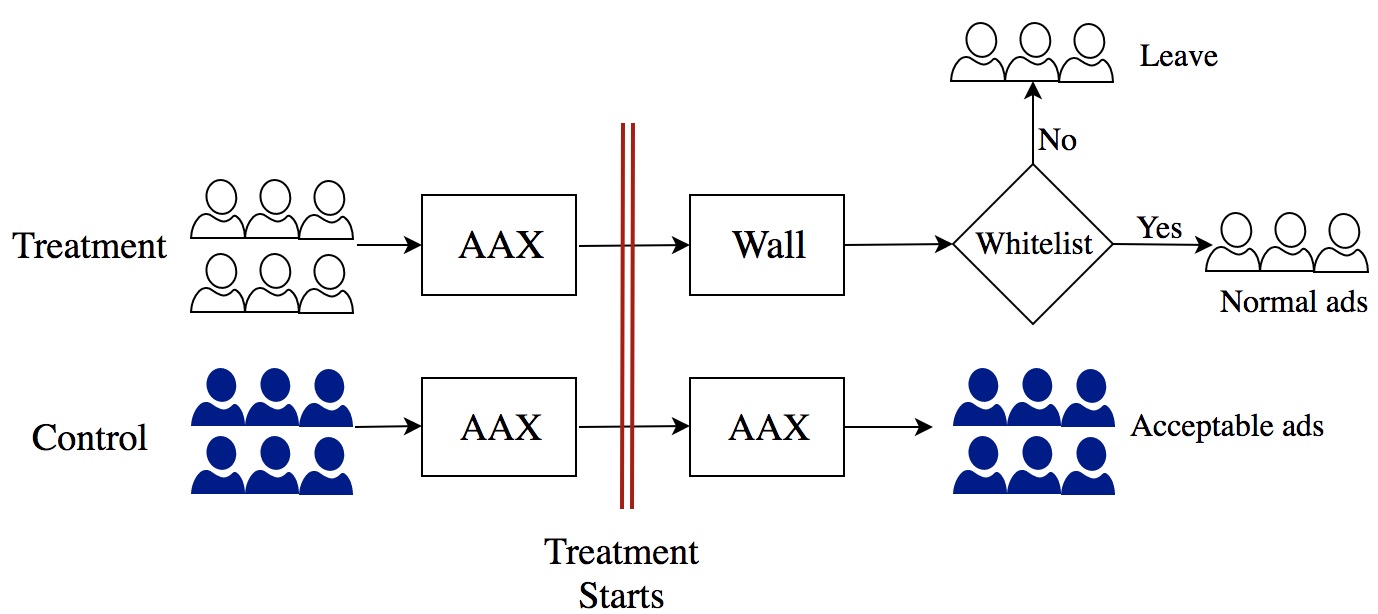, width=0.49\textwidth}
\caption{Illustration of our experiment design}
\label{design}
\end{figure}

Figure~\ref{design} illustrates the experiment design. In the post-treatment period, users in the control group were shown \emph{acceptable ads} and users in the treatment group were wall-blocked and saw an ``Adblock Detected'' message. When facing the Wall strategy, they had to whitelist the web page or the entire web site in order to access the content. Users who did not whitelist left the site.

We utilized JavaScript tracking scripts to detect the existence of ad-blockers. We randomly selected 40,000 ad-blocker users, who were randomly placed into either the control group or the treatment group. The randomization at the user level enables us to track the treatment effect longitudinally, as discussed in Section~\ref{sec:long_study}. 

\section{Data Description} \label{sec:data_descrip}
The dataset contains 40K unique ad-blocker users, equally assigned into either the treatment group or the control group. The data contains a range of user engagement activities and environment measurements such as:

\begin{itemize}
  \item overall and active browser session time
  \item numbers of pages in a session (i.e., pageviews)
  \item hits (i.e., actions) in a session, such as play a video, mouse scroll, or text selection
  \item date and time
  \item geographic location
  \item traffic source (e.g., search engine, social media, or by typing the URL)
  \item system information, e.g., Operating System (OS), browser, screen resolution
\end{itemize}

\begin{wrapfigure}{r}{0.25\textwidth} 
\vspace{-20pt}
  \begin{center}
    \includegraphics[width=0.25\textwidth]{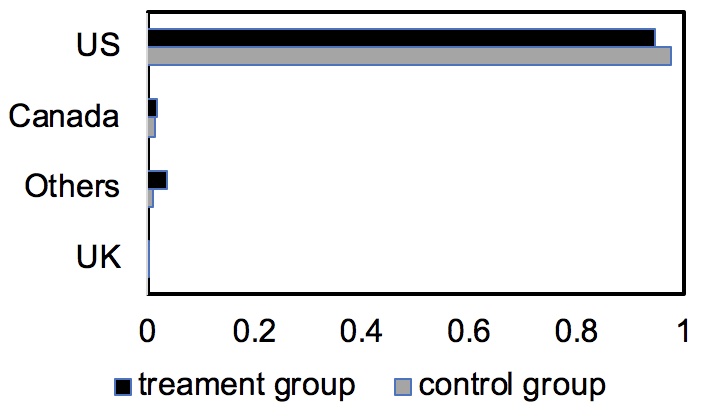}
    \caption{Percentage of users in the study per country}
    \label{country_dis}
  \end{center}
  \vspace{-10pt}
  \vspace{1pt}
\end{wrapfigure}

First, we analyze the characteristics of the ad-blocker users, and the statistics is consistent with the random selection of users between the two groups. The majority of users comes from US because the publisher has high influence on US  audience(Figure~\ref{country_dis}). \sz{Due to the large variety in user attributes, it is difficult to have the distributions of the two groups exactly the same for every attribute. However, the difference in the country distribution of the two groups is small and similar, which confirms the random assignment on users.}

Figure~\ref{browser_os_dis} shows the OS and browser distributions for the users in the dataset. The majority of the ad blocker user visits are from PC operating system, such as Windows and MacOS. The reason is that users are keener to utilize ad-blockers on PCs to avoid annoying ads, since web pages viewed on PCs have more ads and these ads can be intrusive (e.g., video ads). Another reason is that it is easier for users to install ad-blockers on PCs than on mobile devices. For browsers, we find that Chrome is mostly used by ad blocker users, since it is highly popular and offers more ad-blocker software options in its plugin-in store compared to other browsers.

\begin{figure}[ht]
\centering     
\subfigure[OS distribution]{\label{fig:b}\includegraphics[width=0.49\linewidth]{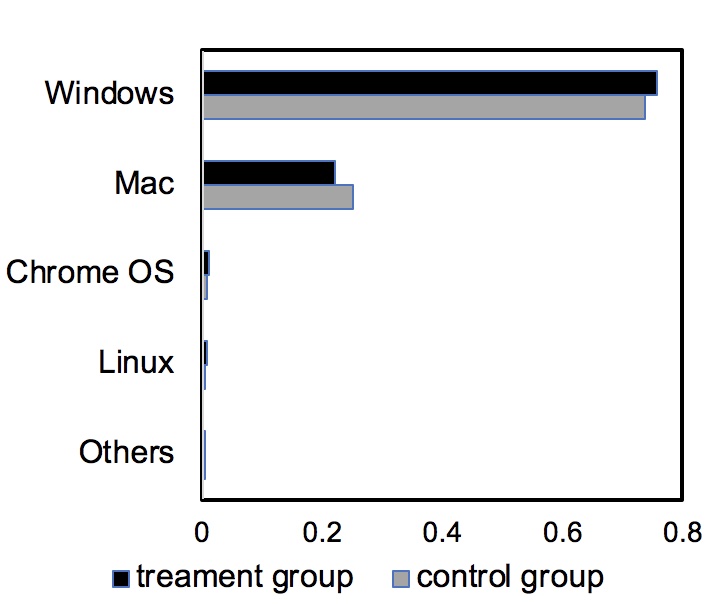}}
\subfigure[Browser distribution]{\label{fig:a}\includegraphics[width=0.45\linewidth]{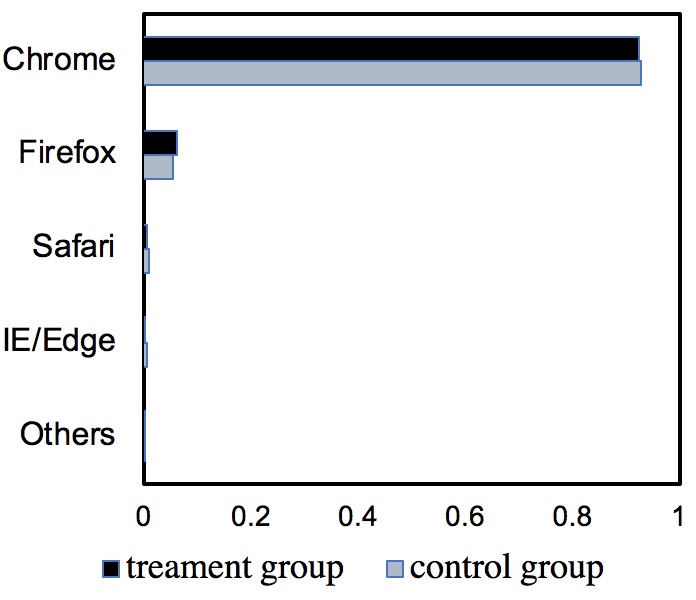}}
\caption{OS and browser distribution for users in the study}
\end{figure} \label{browser_os_dis}

Next, we analyze user behavior across the entire dataset to determine patterns and anomalies. Since we use a real-life dataset, it is inevitable to observe outliers in which the behavior data (e.g., pageviews, hits, or session dwell time) is very high. Extremely large values are probably caused by users leaving the browser open and moving away from the computers. We set a session-level filter threshold for each metric based on the observation of the data distribution to remove such outliers. 

\begin{figure}[ht!]
\centering
\epsfig{file=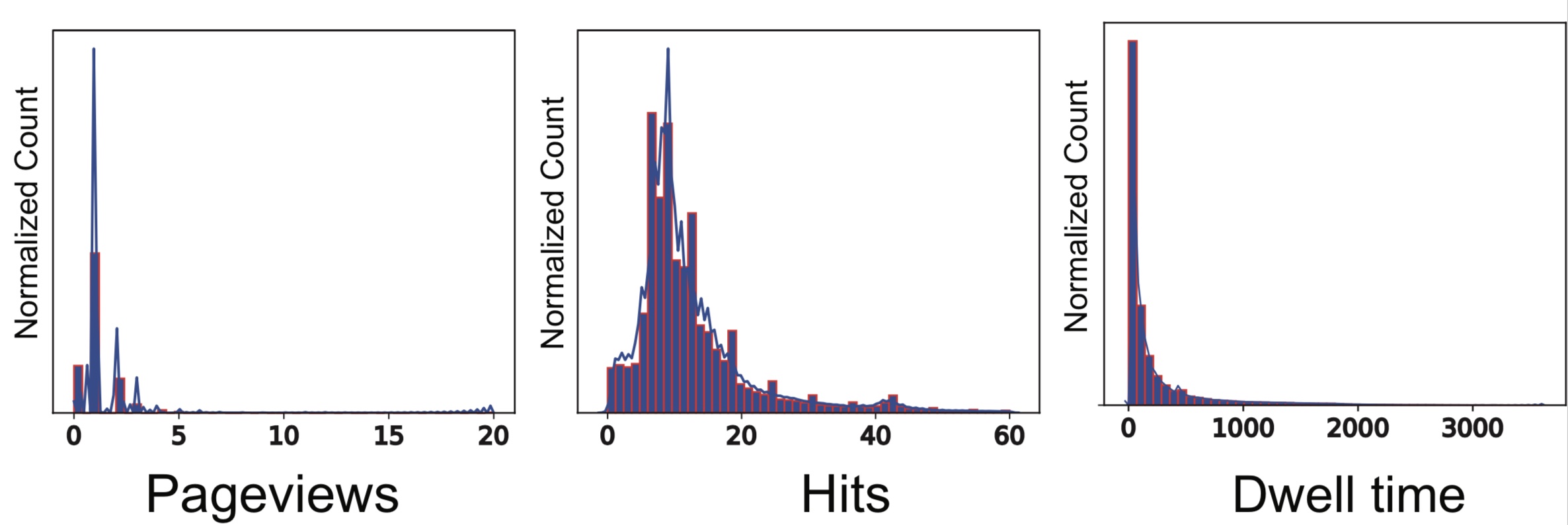, width=0.49\textwidth}
\caption{User behavior ratio (pageviews, hits, and dwell time). The y-axis is the session count normalized by the total number of sessions}
\label{eng_dis}
\end{figure}

The user behavior distribution, after removing outliers, is presented in Figure~\ref{eng_dis}. The distribution fitting curve line for each histogram (i.e., pageviews, hits, and dwell time) is estimated by the kernel density estimation approach. The rightmost values in x-axis in each figure is the filter threshold of outliers. The results show that these user behavior features  are typically  skewed to low values and have a long tail to large values. This is expected because the majority of users tend to have limited interactions with a website in a session. It is also worth mentioning that zero user engagement is recorded if an ad-blocker user choose to leave the website without whitelisting.

\section{Measuring the Impact of the Wall Strategy and Whitelisting} \label{sec:overall_impact}

This section first presents our method for measuring the impact of the wall strategy on user engagement, and further zoom into the whitelist effect, in comparison with the AAX strategy. Then, we analyze the results of applying this method on our data. As suggested by the domain experts from our publisher collaborator, we consider three KPIs (key performance indicators) shown in Table~\ref{dv} to measure user engagement.

\begin{table}[t!]
\caption{KPI Metrics on User Engagement} \label{dv}
\begin{tabular}{p{2cm}p{5cm}}
\hline
\multicolumn{1}{c}{KPI} & \multicolumn{1}{c}{Description}                                      \\ \hline
\#pageviews              & the number of pages viewed in a session                              \\
\#hits                & the number of actions in a session, such as play a video, scroll, selection \\
dwell time            & time spent by a user in a session                                \\ \hline
\end{tabular}
\end{table} 

We use the difference-in-differences (DID) estimation methodology, which is a popular method for estimating average treatment effects (ATE) while controlling for unobservables~\cite{danaher2014effect}. The key underlying assumption of DID is that differences between treatment and control groups would have a common trend in the absence of treatment. It was originally proposed as a ``quasi-experimental'' method to mitigate the effect of extraneous factors and selection bias. The application of DID in our randomized experiment offers robust checks on whether there are group selection bias and extraneous effects. Let us clarify that there is indeed selection bias when measuring the whitelisting effect because the whitelist behavior cannot be randomized in the experiment (i.e., it is decided by the users). Therefore, DID is suitable to measure both the wall strategy and the whitelisting effect in our experiment.

In DID, let $i$ be an ad-blocker user and $Y_{i,j}$ be the engagement outcomes measured by a KPI metric from Table~\ref{dv} that are observed in  session $j$. $T_{i,j}$ is a binary variable regarding the randomly-assigned treatment status, where 1 indicates a user receiving the Wall treatment strategy and 0 indicates a user receiving the AAX control strategy. $t_{i,j}$ is another binary variable regarding the time period, where 1 indicates the time period after the treatment group receives the treatment (i.e., \textbf{post-treatment}) and 0 indicates the time period before the treatment group receives treatment (i.e., \textbf{pre-treatment}). 
The DID can be modeled as linear regression at individual level as follows.

\begin{equation}
Y_{i,j} = \alpha + \beta T_{i,j} + \gamma t_{i,j} + \delta (T_{i,j} * t_{i,j}) + \lambda C_{i,j} + \epsilon_{i,j}
\end{equation}

where $\alpha$, $\beta$, $\gamma$, $\delta$, $\lambda$ are unknown parameters, $C_{i,j}$ is the extraneous factor (i.e., control variables) for user $i$ in session $j$, and $\epsilon_{i,j}$ is a random unobserved ``error'' term. Therefore, the treatment effect is calculated as the difference in the differences of two groups as in the following equation.

\begin{align*}
\hat{\delta_{DD}} &= E[\overline{Y_1}^{T}] - E[\overline{Y_0}^{T}] - (E[\overline{Y_1}^{C}] - E[\overline{Y_0}^{C}]) \\
                 &= \alpha + \beta + \gamma + \delta + \lambda - (\alpha + \beta + \lambda) - (\alpha + \gamma + \lambda - \alpha - \lambda) \\
                 &= (\gamma + \delta) - \gamma \\
                 &= \delta \\
\end{align*}

Here, $\overline{Y_0}^T$ and $\overline{Y_1}^T$ are the sample averages of the behavior outcomes for the treatment group before and after treatment, respectively. $\overline{Y_0}^C$ and $\overline{Y_1}^C$ are the corresponding sample averages of the behavior outcomes for the control group.
The parameter $\delta$ estimates whether the treatment effect is positive or negative, as well as the intensity of the treatment effect. 

A linear model could estimate the treatment effect based on equation~\ref{eq_3}. We add dummy variables in the equation in order to control for the time, the day, and the weekend effect because empirical evidence suggests the user behavior is affected by these factors. hours\_evening and hours\_night are two dummy control variables to indicate whether the visit happens in the evening or night time, compared to during the day time by default.

\begin{equation} \label{eq_3}
\begin{aligned}
    y &= \beta_0 + \beta_1*timeperiod + \beta_2*grouptype + \\
    &\quad \beta_3*hours\_evening + \beta_4*hours\_night + \\
    &\quad \beta_5 * weekend + \beta_6*timeperiod*grouptype + \epsilon
\end{aligned}
\end{equation}

Since a linear model can yield negative predicted values, while our dependent variables should all be non-negative variables, this linear model does not fit well in our study. Inspired by ~\cite{sinha2017anti}, we propose to use a negative binomial (NB) regression for our study. NB regression is based on Poisson regression, which can model non-negative variables. A Poisson regression, however, still possesses one problem for our study. It assumes that the mean and the variance are the same, which may not be not satisfied by the real data. In particular, in our study, the distribution of online user behavior features is typically skewed to the low values and have a long tail to the large values (see Figure~\ref{eng_dis}). The variance is substantially larger than the mean, i.e., over-dispersion. To address the over-dispersion problem caused by highly skewed dependent variables, the NB regression adds a new parameter $\alpha$ in the model. The full NB regression is as follows:

\begin{equation}
\begin{aligned}
P(y|X) = \frac{\Gamma(y+\alpha^{-1})}{y!\Gamma(\alpha^{-1})} (\frac{\alpha^{-1}}{\alpha^{-1} + \mu})^{\alpha^{-1}}(\frac{\mu}{\alpha^{-1} + \mu})^y
\end{aligned}
\end{equation}

$\alpha$ is a positive parameter to represent the extent of over-dispersion auto-fitted by the data. It is solved by the maximum likelihood method. The expected value is $E(y) = \mu$, and the variance is $Var(y) = \mu[1 + \alpha\mu]$, which is a larger than $E(y)$.

\begin{table}[ht!]
\caption{Wall strategy effect on user engagement}
\label{did_2}
\begin{tabular}{llll}
\hline
             & Pageviews                                                   & Hits                                                     & Dwell time                                                  \\ \hline
Constant       & \begin{tabular}[c]{@{}l@{}}0.350***\\ (0.006)\end{tabular}  & \begin{tabular}[c]{@{}l@{}}2.97***\\ (0.003)\end{tabular}  & \begin{tabular}[c]{@{}l@{}}5.793***\\ (0.002)\end{tabular}  \\
hours\_evening & \begin{tabular}[c]{@{}l@{}}-0.024***\\ (0.004)\end{tabular} & \begin{tabular}[c]{@{}l@{}}-0.004\\ (0.002)\end{tabular}    & \begin{tabular}[c]{@{}l@{}}-0.079***\\ (0.002)\end{tabular} \\
hours\_night   & \begin{tabular}[c]{@{}l@{}}-0.015*\\ (0.006)\end{tabular}   & \begin{tabular}[c]{@{}l@{}}-0.015***\\ (0.003)\end{tabular} & \begin{tabular}[c]{@{}l@{}}-0.098***\\ (0.003)\end{tabular} \\
group      & \begin{tabular}[c]{@{}l@{}}0.052***\\ (0.006)\end{tabular}  & \begin{tabular}[c]{@{}l@{}}0.014***\\ (0.003)\end{tabular}  & \begin{tabular}[c]{@{}l@{}}0.041***\\ (0.002)\end{tabular}  \\
period     & \begin{tabular}[c]{@{}l@{}}0.044***\\ (0.005)\end{tabular}  & \begin{tabular}[c]{@{}l@{}}0.014***\\ (0.002)\end{tabular}  & \begin{tabular}[c]{@{}l@{}}0.001\\ (0.002)\end{tabular}     \\
weekend        & \begin{tabular}[c]{@{}l@{}}0.032***\\ (0.005)\end{tabular}  & \begin{tabular}[c]{@{}l@{}}-0.032***\\ (0.002)\end{tabular} & \begin{tabular}[c]{@{}l@{}}-0.018***\\ (0.002)\end{tabular} \\
\textbf{group*period}            & \textbf{\begin{tabular}[c]{@{}l@{}}-0.215***\\ (0.008)\end{tabular}} & \textbf{\begin{tabular}[c]{@{}l@{}}-0.456***\\ (0.004)\end{tabular}} & \textbf{\begin{tabular}[c]{@{}l@{}}-0.262***\\ (0.003)\end{tabular}} \\ \hline
\end{tabular}
    \begin{tablenotes}
      \small
      \item Note. Cluster-robust standard errors in parentheses.
      \item *p < 0.1; **p < 0.05; ***p < 0.01.
    \end{tablenotes}
\end{table}

\textbf{Wall Strategy Effect}: We start with the session level measurement. The results of the NB regression are in Table~\ref{did_2}. The Wall strategy has a negative effect on user engagement according to $\delta$, i.e., the coefficient parameter $\beta_6$ in equation~\ref{eq_3}.
The effect includes a statistically significant decrease of $e^{-0.215}-1=-19.3\%$ on the number of pageviews, $e^{-0.456}-1=-36.6\%$ on the number of hits, and $e^{-0.262}-1=-23.0\%$ on the session dwell time, compared to the AAX strategy. The results are as expected because some users in the treatment group choose not to whitelist, and thus they are denied access to page content, consequently, resulting in less engagement for the treatment population. But when examining the coefficients ahead of \emph{group}, we find it is statistically significant. We think it is due to the large variety of unobserved user attributes instead of problems with our randomization. Also, compared to the true treatment effect, the magnitude of group variable is small and inconsequential.

\textbf{Zoom Into Users Who Whitelist:} 
We next zoom into the users who choose to whitelist in the treatment group.  We compare engagement behaviors in the whitelisted sessions of the treatment group with the control group where users have AAX sessions. As can be seen in Table~\ref{did_3}, the whitelist behavior has a statistically significant positive effect on user engagement. The intuition is that  whitelist behavior indicates that the ad-blocker user has higher interest in the intended article, and thus is more likely to  spend more time and interact more with the website than the users in the control group.

\begin{table}[ht!]
\caption{Whitelist effect on user engagement}
\label{did_3}
\begin{tabular}{llll}
\hline
             & Pageviews                                                   & Hits                                                     & Dwell Time                                                  \\ \hline
Constant       & \begin{tabular}[c]{@{}l@{}}0.365***\\ (0.005)\end{tabular}  & \begin{tabular}[c]{@{}l@{}}2.967***\\ (0.003)\end{tabular}  & \begin{tabular}[c]{@{}l@{}}5.799***\\ (0.002)\end{tabular}  \\
hours\_evening & \begin{tabular}[c]{@{}l@{}}-0.028***\\ (0.003)\end{tabular} & \begin{tabular}[c]{@{}l@{}}-0.024***\\ (0.002)\end{tabular} & \begin{tabular}[c]{@{}l@{}}-0.102***\\ (0.001)\end{tabular} \\
hours\_night   & \begin{tabular}[c]{@{}l@{}}-0.024***\\ (0.005)\end{tabular} & \begin{tabular}[c]{@{}l@{}}-0.023***\\ (0.002)\end{tabular} & \begin{tabular}[c]{@{}l@{}}-0.109***\\ (0.002)\end{tabular} \\
group      & \begin{tabular}[c]{@{}l@{}}0.052***\\ (0.006)\end{tabular}  & \begin{tabular}[c]{@{}l@{}}0.015***\\ (0.003)\end{tabular}  & \begin{tabular}[c]{@{}l@{}}0.041***\\ (0.002)\end{tabular}  \\
period     & \begin{tabular}[c]{@{}l@{}}-0.026***\\ (0.004)\end{tabular} & \begin{tabular}[c]{@{}l@{}}-0.365***\\ (0.002)\end{tabular} & \begin{tabular}[c]{@{}l@{}}-0.352***\\ (0.002)\end{tabular} \\
weekend        & \begin{tabular}[c]{@{}l@{}}0.019***\\ (0.004)\end{tabular}  & \begin{tabular}[c]{@{}l@{}}-0.010***\\ (0.002)\end{tabular} & \begin{tabular}[c]{@{}l@{}}-0.010***\\ (0.002)\end{tabular} \\
\textbf{group*period}            & \textbf{\begin{tabular}[c]{@{}l@{}}0.218***\\ (0.010)\end{tabular}}  & \textbf{\begin{tabular}[c]{@{}l@{}}0.128***\\ (0.006)\end{tabular}}  & \textbf{\begin{tabular}[c]{@{}l@{}}0.158***\\ (0.005)\end{tabular}}  \\ \hline
\end{tabular}
    \begin{tablenotes}
      \small
      \item Note. Cluster-robust standard errors in parentheses.
      \item *p < 0.1; **p < 0.05; ***p < 0.01.
    \end{tablenotes}
\end{table}

\section{Analyzing the Wall Effect on User Groups} \label{sec:cluster}

\begin{wrapfigure}{r}{0.25\textwidth} 
\vspace{-9pt}
  \begin{center}
    \includegraphics[width=0.25\textwidth]{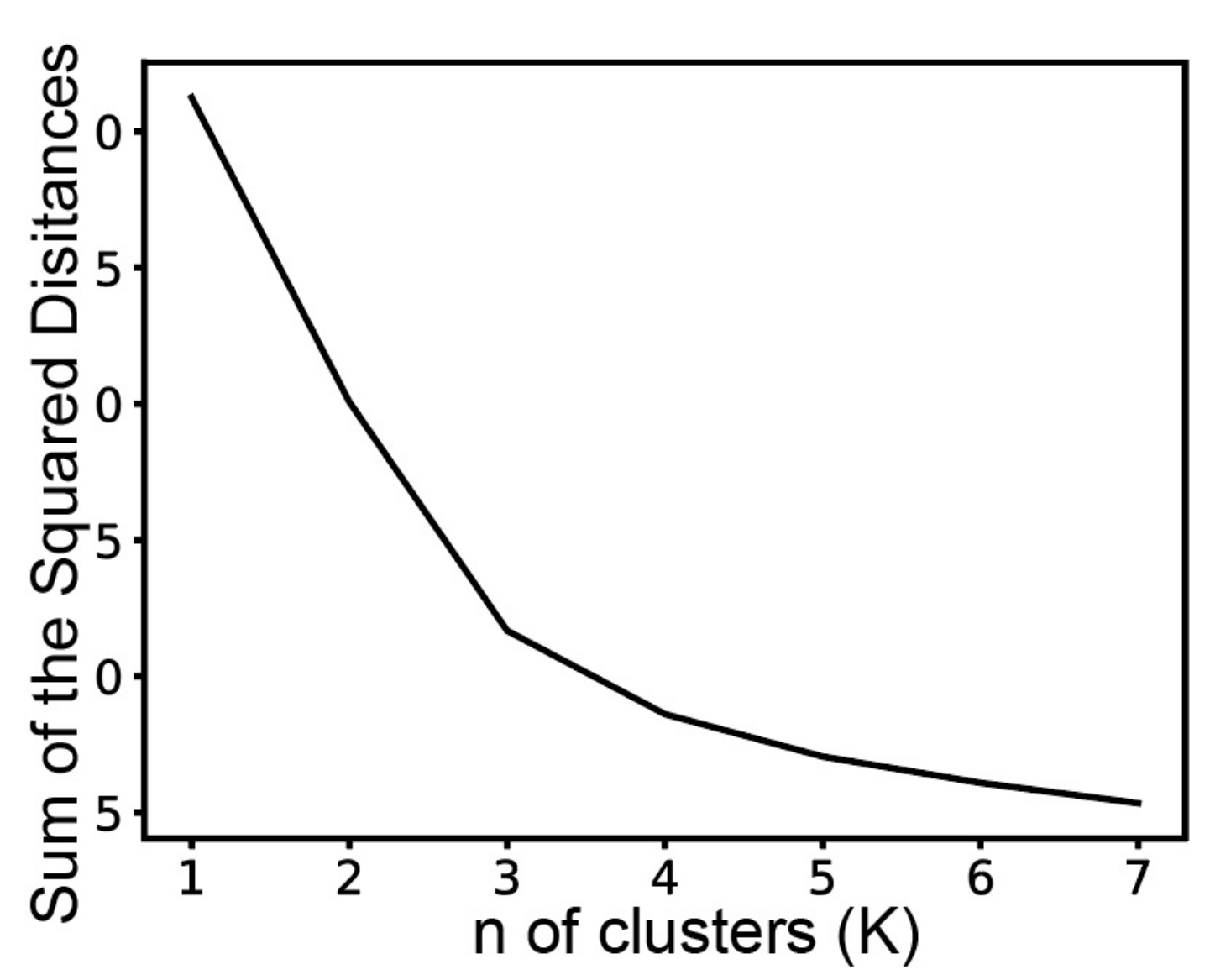}
    \caption{Choosing the number of clusters}
    \label{cluster}
  \end{center}
  \vspace{-16pt}
\end{wrapfigure}

This section measures the impact of the Wall strategy on users with different characteristics. User loyalty is a major characteristic impacting a user's behavior, and it is represented by her engagement with the website. Therefore, we propose to cluster users based on their engagement level, observed in the pre-treatment period. In other words, user characteristics are identified before the treatment starts. 

The clustering features include the total number of sessions, the numbers of pageviews and hits, and the dwell time. 
The K-Means method is used for clustering, and the Euclidean distance is used to measure the similarity between users. Figure~\ref{cluster} shows the sum of Euclidean distances for each user to its nearest centroid (y-axis), with varying numbers of clusters (x-axis). The shape of the fitting curve suggests the number of cluster $K=3$ is a good choice because it is at the elbow of the curve. 

\begin{wrapfigure}{r}{0.25\textwidth}
\vspace{-15pt}
  \begin{center}
    \includegraphics[width=0.25\textwidth]{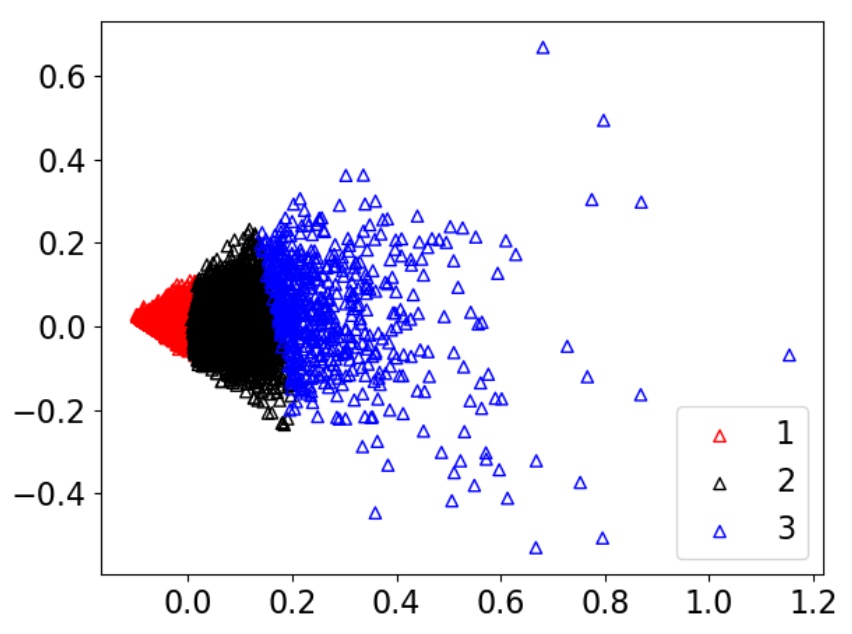}
    \caption{Visualization of user clustering results by PCA; Axes are latent dimensions}
    \label{pca_2d}
  \end{center}
  \vspace{-20pt}
  \vspace{1pt}
\end{wrapfigure}

The user cluster results are described in Table~\ref{cluster_discribe}. The user engagement increases from group 1 to group 3, with group 1 consisting of low-engaged users, group 2 consisting of medium-engaged users, and group 3 consisting of high-engaged users. It shows that the majority of users are low engaged. We use the principal component analysis method to reduce the engagement features into two dimensions and visualize the clusters in Figure~\ref{pca_2d}. The figure shows the three groups of users are clustered in different areas in space. The low-engaged users are crowded together in one small area due to few sessions and limited engagement. The high-engaged users spread into a much larger area due to more frequent visits and higher variant engagement, and medium-engaged users are in the middle.

In order to measure the cluster-level impact of the Wall strategy, we use the coarsened exact matching method to match individual users in the treatment and control groups on a one-to-one basis to make sure that the samples are balanced and the users are similar to each other. The reason is that user engagement tends to exhibit the ``regression-towards-the-mean'' (RTM) phenomenon~\cite{barnett2004regression}. 
In order to avoid the RTM influence and sample bias, we design a matching procedure, as illustrated in Algorithm~\ref{alg:generator}. We utilize the same engagement features as in the clustering. Euclidean distance is selected to measure the user similarity. If there is no similar user in the control group (i.e., exceeds the threshold), we will discard the corresponding user in the treatment group. Overall, 99\% of users in the treatment groups are matched to users in the control group.

\begin{algorithm}
\caption{User Matching}
\label{alg:generator}
\SetKwProg{generate}{Function}{}{end}

Map store=new Map(user\_id, user\_id)\;
\generate{Matcher}{
     \ForAll{user $i$ in one cluster of treatment group}{
        \ForAll{user $j$ in control group $C$}{calculate the distance between $i$ and $j$, find the closest user $j^{*}$ for $i$;}
        \If{distance($i$, $j^{*}$) < threshold}{add ($i$, $j^{*}$) into store; delete $j^*$ from $C$;}
     }
     return store;
     
}
\end{algorithm}

\begin{table}[ht!]
\caption{Descriptive Analysis of User Clusters}
\begin{tabular}{llllll}
\hline
Group & \% Users & \% Visits & \begin{tabular}[c]{@{}l@{}}Average\\ session\\ \#pageviews\end{tabular} & \begin{tabular}[c]{@{}l@{}}Average\\ session\\ \#hits \end{tabular} & \begin{tabular}[c]{@{}l@{}}Session \\ dwell \\ time\end{tabular} \\ \hline
1        & 85.76    & 55.76     & 1.37                                                                    & 17.3                                                                  & 218.6                                                         \\
2        & 12.88    & 32.96     & 1.53                                                                    & 20.4                                                                 & 414.7                                                          \\
3        & 1.36     & 11.28      & 1.85                                                                    & 21.6                                                                  & 528.6                                                        \\ \hline
\end{tabular} \label{cluster_discribe}
\end{table}

The DID method is utilized to measure the Wall strategy effect per cluster and the results are shown in Table~\ref{cluster_result}. We notice that the coefficients of \textit{group} are close to 0, and they are not statistically significant.
This validates the effectiveness of our user matching procedure. It avoids the selection bias per cluster, and it indeed matches similar users in the treatment and control groups. 

\begin{table*}[htb!]
\caption{Effect of The Wall Strategy per User Cluster}
\begin{tabular}{llllllllll}
\hline
                      & \multicolumn{3}{c}{cluster 1}                                                                                                                                                                                     & \multicolumn{3}{c}{cluster 2}                                                                                                                                                                                    & \multicolumn{3}{c}{cluster 3}                                                                                                                                                                             \\ \hline
                      & Pageviews                                                            & Hits                                                                 & Dwell time                                                          & Pageviews                                                          & Hits                                                                 & Dwell Time                                                           & Pageviews                                                         & Hits                                                              & Dwell time                                                        \\ \hline
constant              & \begin{tabular}[c]{@{}l@{}}0.365***\\ (0.011)\end{tabular}           & \begin{tabular}[c]{@{}l@{}}2.898***\\ (0.008)\end{tabular}           & \begin{tabular}[c]{@{}l@{}}5.490***\\ (0.008)\end{tabular}          & \begin{tabular}[c]{@{}l@{}}0.525***\\ (0.038)\end{tabular}         & \begin{tabular}[c]{@{}l@{}}3.202***\\ (0.031)\end{tabular}           & \begin{tabular}[c]{@{}l@{}}6.220***\\ (0.030)\end{tabular}           & \begin{tabular}[c]{@{}l@{}}0.671***\\ (0.085)\end{tabular}        & \begin{tabular}[c]{@{}l@{}}3.218***\\ (0.071)\end{tabular}        & \begin{tabular}[c]{@{}l@{}}6.413***\\ (0.069)\end{tabular}        \\
period                & \begin{tabular}[c]{@{}l@{}}-0.030**\\ (0.015)\end{tabular}           & \begin{tabular}[c]{@{}l@{}}-0.294***\\ (0.012)\end{tabular}          & \begin{tabular}[c]{@{}l@{}}-0.196***\\ (0.012)\end{tabular}         & \begin{tabular}[c]{@{}l@{}}-0.160***\\ (0.055)\end{tabular}        & \begin{tabular}[c]{@{}l@{}}-0.462***\\ (0.044)\end{tabular}          & \begin{tabular}[c]{@{}l@{}}-0.535***\\ (0.043)\end{tabular}          & \begin{tabular}[c]{@{}l@{}}-0.231*\\ (0.123)\end{tabular}         & \begin{tabular}[c]{@{}l@{}}-0.403***\\ (0.100)\end{tabular}       & \begin{tabular}[c]{@{}l@{}}-0.439***\\ (0.098)\end{tabular}       \\
group                 & \begin{tabular}[c]{@{}l@{}}0.022\\ (0.015)\end{tabular}              & \begin{tabular}[c]{@{}l@{}}0.001\\ (0.001)\end{tabular}              & \begin{tabular}[c]{@{}l@{}}0.006\\ (0.012)\end{tabular}             & \begin{tabular}[c]{@{}l@{}}0.028\\ (0.054)\end{tabular}            & \begin{tabular}[c]{@{}l@{}}-0.005\\ (0.044)\end{tabular}             & \begin{tabular}[c]{@{}l@{}}0.007\\ (0.043)\end{tabular}              & \begin{tabular}[c]{@{}l@{}}0.046\\ (0.120)\end{tabular}           & \begin{tabular}[c]{@{}l@{}}-0.008\\ (0.102)\end{tabular}          & \begin{tabular}[c]{@{}l@{}}0.022\\ (0.098)\end{tabular}           \\
\textbf{group*period} & \textbf{\begin{tabular}[c]{@{}l@{}}-0.258***\\ (0.022)\end{tabular}} & \textbf{\begin{tabular}[c]{@{}l@{}}-0.162***\\ (0.017)\end{tabular}} & \textbf{\begin{tabular}[c]{@{}l@{}}-0.034**\\ (0.016)\end{tabular}} & \textbf{\begin{tabular}[c]{@{}l@{}}-0.114*\\ (0.078)\end{tabular}} & \textbf{\begin{tabular}[c]{@{}l@{}}-0.226***\\ (0.062)\end{tabular}} & \textbf{\begin{tabular}[c]{@{}l@{}}-0.114***\\ (0.044)\end{tabular}} & \textbf{\begin{tabular}[c]{@{}l@{}}-0.020\\ (0.173)\end{tabular}} & \textbf{\begin{tabular}[c]{@{}l@{}}-0.021\\ (0.042)\end{tabular}} & \textbf{\begin{tabular}[c]{@{}l@{}}-0.079\\ (0.138)\end{tabular}} \\ \hline
\end{tabular} \label{cluster_result}
    \begin{tablenotes}
      \small
      \item Note. Cluster-robust standard errors in parentheses. *p < 0.1; **p < 0.05; ***p < 0.01.
    \end{tablenotes}
\end{table*}

As shown in Table~\ref{cluster_result}, the Wall strategy does not have statistically significant effect on high-engaged users. This is expected because these users are loyal, and they would view the pages no matter what strategy is used. For medium-engaged users, the Wall strategy hurts the most on their interactions with the pages (hits and dwell time), but less on pageviews. Medium-engaged users still need to access the pages, but their activities are largely weakened by the intrusiveness of the Wall strategy or the annoying ads after whitelisting (i.e., the medium-engaged users in the control AAX strategy see less annoying ads).

For low-engaged users, the Wall strategy has a large negative effect on pageviews, but relatively low impact on hits and dwell time. This is because low-engaged users have no loyalty to the website and their whitelist decisions are  driven by their interest on the intended page. Thus, a significant amount of low-engaged users will refuse to whitelist and leave, resulting in a large decrease on pageviews. On the other hand, their original base of hits and dwell time are the lowest, and they cannot decrease much after the treatment. Therefore, the effect of the Wall strategy on hits and dwell time for this user group is smaller than the one for the medium-engaged user group. From a publisher's point of view, the number of pageviews is more important than hits and dwell time because of the popular cost-per-view business charging model to advertisers. Also, since the majority of users are low-engaged users, the revenue of the publisher is expected to suffer a lot when using the Wall strategy.

\section{Long-Term Study} \label{sec:long_study}

\begin{wrapfigure}{r}{0.3\textwidth} 
\vspace{-20pt}
  \begin{center}
    \includegraphics[width=0.3\textwidth]{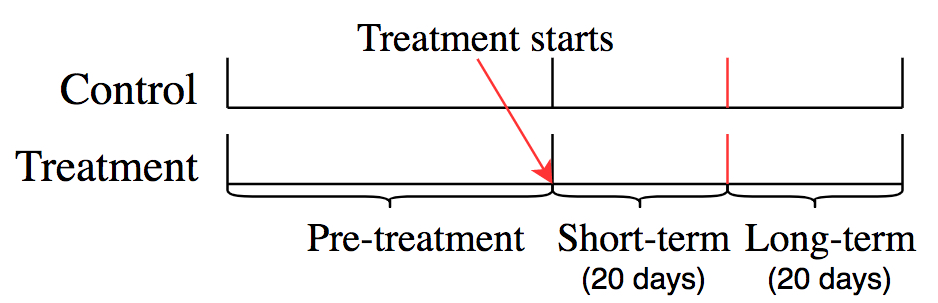}
    \caption{Short-term and long-term post-treatment measurements}
    \label{seperate}
  \end{center}
  \vspace{-15pt}
  \vspace{1pt}
\end{wrapfigure} 

Next, we study the effect of  the Wall strategy on user engagement over time. We separate our post-treatment period into two equal sub-periods (20 days interval in each sub-period),  as illustrated in Figure~\ref{seperate}. We refer to the first 20 days as short-term and to the next 20 days as long-term.

We consider three aspects to measure the long-term effect of the Wall strategy on user engagement: the frequency of revisiting, whitelist ratio, and  user engagement per session. First, we compare the number of visits per user as well as the session-level whitelist ratio for short-term and long-term periods in the treatment group. 

\begin{figure}[ht]
\centering     
\subfigure[\#Visits and whitelist ratio ]{\label{fig:short_long1}\includegraphics[width=0.43\linewidth]{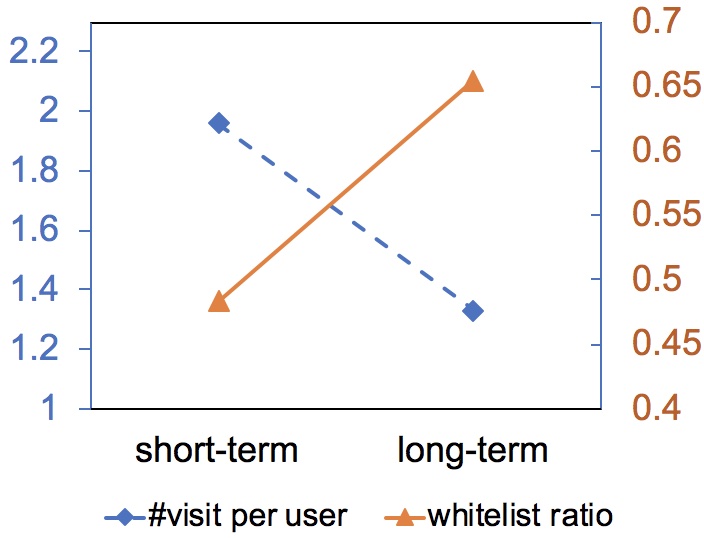}}
\subfigure[Next visit]{\label{fig:survival}\includegraphics[width=0.49\linewidth]{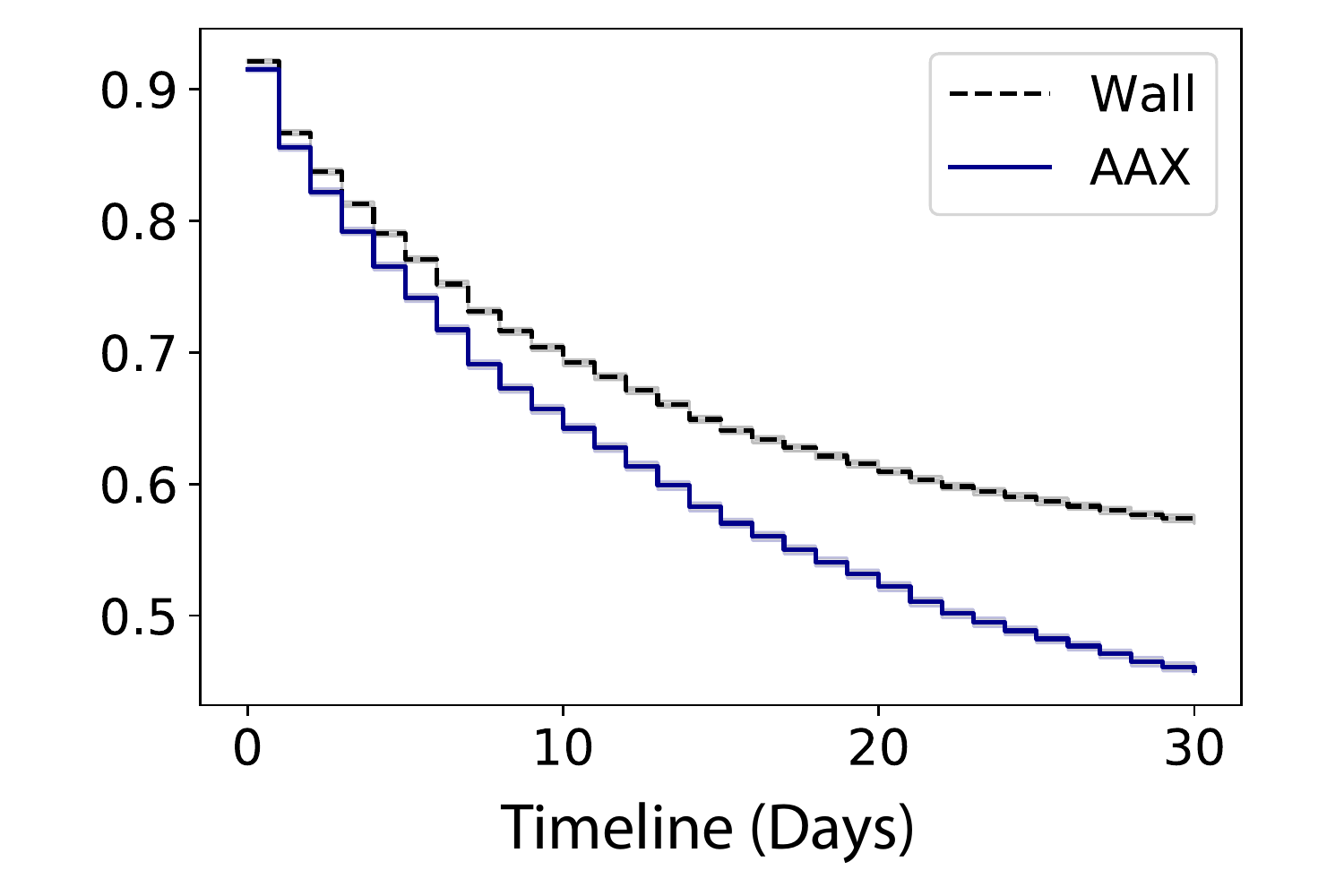}}
\caption{Comparison between short-term and long-term visit behavior}
\end{figure} \label{short_long}

\begin{wrapfigure}{r}{0.22\textwidth} 
\vspace{-10pt}
  \begin{center}
    \includegraphics[width=0.22\textwidth]{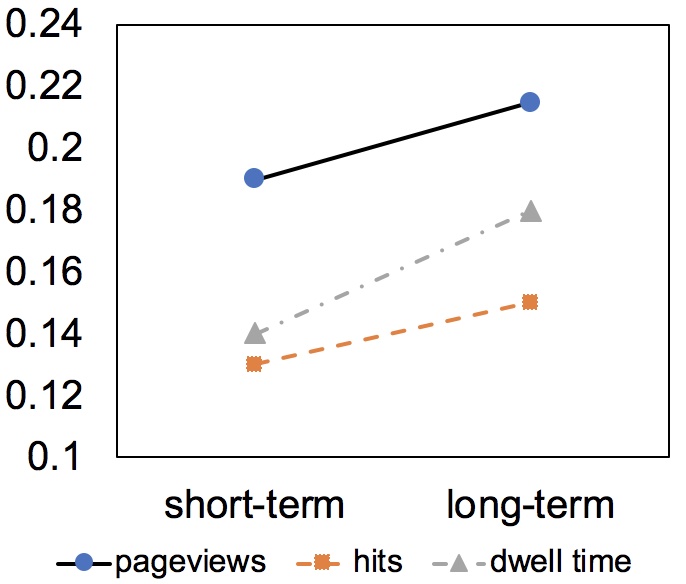}
    \caption{Engagement in a session}
    \label{whitelist_long}
  \end{center}
  \vspace{-10pt}
  \vspace{1pt}
\end{wrapfigure} 

As shown in Figure~\ref{fig:short_long1}, the average number of visits per user within a subperiod decreases from 1.96 to 1.33, which indicates that fewer revisits happen over time. To better examine the effect of the Wall strategy on revisiting, inspired by~\cite{stoolmiller2006modeling}, we utilize the Kaplan-Meier estimator to fit the survival curves of revisits. We consider the duration gap between the first initial Wall treatment and the next visit. The results are shown in Figure~\ref{fig:survival}, in which X axis is the timeline in terms of days, and the Y axis is the percentage of no revisits. The figure shows the black dashed line (the Wall strategy) is above the the blue solid line (the AAX strategy), indicating that the Wall strategy postpones the next revisit of the same user in the treatment group.  Quantitatively, we find that the Wall strategy causes a 20.5\% increase of the visit duration gap. The reason is probably that the ad-blocker users feel disturbed when facing the Wall strategy, and they are less willing to come back.

We also observe that the whitelisted-session ratio increases from 48.3\% to 65.5\%. The reason is that, with the Wall strategy, loyal users are likelier to whitelist gradually over time. On the other hand, the ad-blocker users who refuse to whitelist previously would probably not come back again.

Finally, we measure the user engagement behavior for each session over time, where we consider only the whitelist sessions. Similar to the method presented in Section~\ref{sec:overall_impact}, we use the DID method to control the extraneous variables. As shown in Figure~\ref{whitelist_long}, there is a slight increase in the engagement behavior in a whitelist session over time. This indicates that users get accustomed to the Wall strategy and the annoying ads over time. It also shows the Wall strategy effect is stronger in the short-term, but its negative effect is reduced gradually over time.

\section{Discussions} \label{sec:conclusion}

In this paper, we conduct a randomized field experiment on two counter-ad-blocking strategies, benefiting from collaboration with Forbes Media, a major US media company. Our analysis shows that the Wall strategy has indeed a filtering effect on high-engaged users. They have strong loyalty to websites, and are more likely to whitelist. Therefore, we do not recommend the Wall strategy to  publishers unless they have a large portion of loyal users.

If a publisher indeed wants to adopt the Wall strategy, the problem is how to convert casual users to high-engaged users, since casual users are more likely to leave forever when facing the Wall strategy. Our suggestion is to allow new users to bypass the Wall in order to strengthen their attachment to the website. The Wall can then be shown later, after noticing a significant increase in their engagement. Nevertheless, future research can expand this work by designing dynamic wall blocking strategies using machine learning methods to optimize user conversion to high-engaged users.

\section*{Acknowledgement}
\sz{This work is partially supported by NSF under grant No. DGE 1565478, and by the Leir Foundation. Any opinions, findings, and conclusions expressed in this material are those of the authors and do not necessarily reflect the views of the funding agencies.}

\bibliographystyle{ACM-Reference-Format}

\bibliography{bibliography}

\end{document}